\documentclass[epj]{svjour}
\usepackage{graphics}
\begin{document}
\title{The role of the gluonic $gg\leftrightarrow ggg$ interactions
in early thermalization in ultrarelativistic heavy-ion collisions}
\author{Zhe Xu\inst{1}
\thanks{\emph{xu@th.physik.uni-frankfurt.de}}%
\and Carsten Greiner\inst{1}
}                     
%
%
\institute{Institut f\"ur Theoretische Physik, Johann Wolfgang
Goethe-Universit\"at Frankfurt, Germany}
\date{Received: date / Revised version: date}
%
\titlerunning{The role of the gluonic $gg\leftrightarrow ggg$ interactions
in early thermalization}

\abstract{
We ``quantify'' the role of elastic as well as inelastic
$gg \leftrightarrow ggg$ pQCD processes in kinetic equilibration 
within a pQCD inspired parton cascade. The contributions of different
processes to kinetic equilibration are manifested
by the {\em transport collision rates}. We find that in a central Au+Au 
collision at RHIC energy pQCD Bremstrahlung processes are much more
efficient for momentum isotropization compared to elastic scatterings.
For the parameters chosen the ratio of their transport collision rates 
amounts to $5:1$.
\PACS{
      {05.60.-k}{Transport processes}    \and
      {25.75.-q}{Relativistic heavy-ion collisions}   \and
      {24.10.Lx}{Monte Carlo simulations}
     } 
} 
\maketitle
\section{Introduction}
\label{intro}
It was speculated that a strongly coupled quark-gluon plasma (sQGP)
\cite{Guyl} is formed in Au+Au collisions at RHIC. This
strong coupling, or strong interaction, makes the QGP to a fluid
with very small viscosity. However, how strong the coupling must be
in order to generate a quasi-ideal fluid is an open question.

Recently we have developed an on-shell parton cascade including elastic 
as well as inelastic $gg \leftrightarrow ggg$ pQCD processes to study 
the issue of thermalization \cite{Xu1}. Although the total cross section 
of the pQCD scatterings is a few $mb$, it is enough to drive the
system into thermal equilibrium and also to generate a large elliptic
flow $v_2$ in noncentral Au+Au collisions \cite{Xu2}.

Since with elastic pQCD scatterings alone no thermalization is expected 
to be achieved \cite{Xu1,serreau}, the inelastic $gg \leftrightarrow ggg$ 
processes seem to play a leading role in early thermalization, although 
for the parameters chosen in \cite{Xu1} the cross section of $gg\to ggg$ 
collisions is a factor of 2 smaller than that of elastic scatterings. 
In order to understand this a {\em transport cross section},
\begin{equation}
\label{tcs}
\sigma^{tr.}=\int d\theta \frac{d\sigma}{d\theta} sin^2\theta\,,
\end{equation}
was introduced as a pertinent quantity measuring the contributions
of different collision processes to kinetic equilibration \cite{dg},
since large-angle collisions should contribute more to momentum 
isotropization. The results (see Fig. 48 in \cite{Xu1}) showed that even
due to the almost isotropic distribution of the collision angles in
inelastic collisions its transport cross section is only the same as that
of elastic scatterings. Therefore one cannot understand why the inclusion 
of the inelastic processes brings so much to kinetic equilibration. 
At first sight it seems to be an unsolvable problem. On the other hand,
however, there is no reason to believe that momentum isotropization should
relate to the angular distribution by means of the transport cross 
section and not by another formula. The concept of the transport
cross section may be more intuitively than mathematically
correct. In this work we will find a mathematically correct way to
quantify the contribution of different processes to thermal 
equilibration and compare them with each other. The core issue is
the {\em transport collision rate}. With this quantity we ``quantify''
the role of different collision processes in kinetic equilibration.

\section{Parton cascade}
\label{sec:1}
The buildup of the parton cascade is based on the stochastic interpretation
of the transition rate. This guarantees detailed balance,
which is, by contrast, difficult when using the geometrical concept 
of the cross section \cite{molnar}, especially for multiple scatterings like 
$ggg \to gg$. The particular feature of the numerical implementation in
the parton cascade is the subdivision of space into small cell units. In 
cells the transition probabilities are evaluated for random sampling whether
a particular scattering occurs or not. The smaller the cells, the more 
locally transitions will be realized.

The three-body gluonic interactions are described by the matrix element
\cite{gb}
\begin{eqnarray}
\label{m23}
| {\cal M}_{gg \to ggg} |^2 &=& \frac{9 g^4}{2} 
\frac{s^2}{({\bf q}_{\perp}^2+m_D^2)^2}\,
 \frac{12 g^2 {\bf q}_{\perp}^2}
{{\bf k}_{\perp}^2 [({\bf k}_{\perp}-{\bf q}_{\perp})^2+m_D^2]} \times
\nonumber \\
&&\times \, \Theta(k_{\perp}\Lambda_g-\cosh y)\,.
\end{eqnarray} 
The suppression of the radiation of soft gluons due to the 
Landau-Pomeranchuk-Migdal (LPM) effect \cite{biro,wong,Xu1},
which is expressed via the step function in Eq. (\ref{m23}),
is modeled by the consideration that the time of the emission,
$\sim \frac{1}{k_{\perp}} \cosh y$, should be smaller than the time
interval between two scatterings or equivalently  the gluon mean free
path $\Lambda_g$. This leads to a lower cutoff for $k_{\perp}$ and
to a decrease of the total cross section.

In this work we simulate the time evolution of gluons produced in a
central Au+Au collision at RHIC energy. The initial gluons are taken as 
minijets with transverse momentum being greater than $1.4$ GeV, which are 
produced via semi-hard nucleon-nucleon collisions. Using the Glauber geometry
the gluon number is initially about $700$ per momentum rapidity. These gluons 
take about $60\%$ of the total energy entered in the collision. Choosing 
such an initial condition and performing a simulation including
Bremsstrahlung processes we obtain $dE_T/dy$ about $640$ GeV at midrapidity 
at a final time of $5$ fm/c, at which the energy density of gluons
decreases to the critical value of $1$ GeV/$\mbox{fm}^3$. The value of 
$dE_T/dy$ obtained from the simulation is comparable with that from the 
experimental measurements at RHIC.

We concentrate on the central region: $0 < x_T < 1.5$ fm and 
$-0.2 < \eta < 0.2$, where $\eta$ denotes the space-time rapidity. 
Results which will be shown below are obtained in this region by 
ensemble average.

\begin{figure}[h]
\begin{center}
\resizebox{0.4\textwidth}{!}{
  \includegraphics{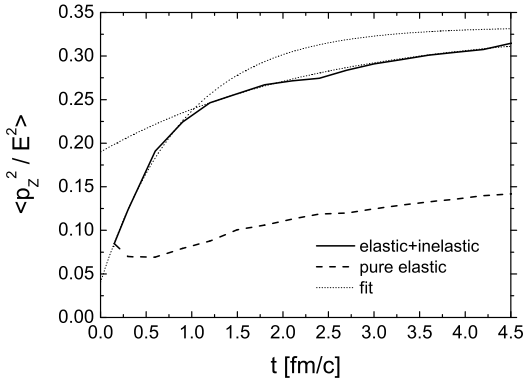}
}
\end{center}
\caption{Momentum anisotropy.}
\label{fig:1}
\end{figure}

The importance of including inelastic pQCD $gg\leftrightarrow ggg$
processes to momentum isotropization is clearly demonstrated in 
Fig. \ref{fig:1}, where the time evolution of the averaged momentum 
anisotropy, $<p_Z^2/E^2>$, is depicted.
$p_Z$ and $E$ are, respectively, longitudinal momentum and energy of
a gluon. The average is computed over all gluons in the central region.
As comparison we have also performed a simulation with pure elastic 
processes starting with the same initial conditions.
From Fig. \ref{fig:1} we see that while the gluon system is still far
from kinetic equilibrium in the simulation with pure elastic scatterings
(dashed curve), the momentum anisotropy relaxes to the value at 
equilibrium, $1/3$, in the simulation including inelastic processes
(solid curve). 

We fit the time evolution of the momentum anisotropy 
using the standard relaxation formula
\begin{equation}
\label{fit}
F(t)=\frac{1}{3}+ \left ( \left < \frac{p_Z^2}{E^2} \right >(t_0)
-\frac{1}{3} \right ) \exp\, (-\frac{t-t_0}{\theta(t_0)})\,.
\end{equation}
For simplicity we label now the momentum anisotropy by $Q:=<p_Z^2/E^2>$.
$F(t)$ is only equal to $Q(t)$ at $t=t_0$. For fixed $t_0$ the
relaxation time $\theta$ is a constant with respect to $t$.
Such a fit can be done at every time point $t_0$. The two thin 
dotted curves in Fig. \ref{fig:1} are fits with $\theta=0.9$ fm/c at 
$t_0=0.3$ fm/c and $\theta=2.4$ fm/c at $t_0=1.2$ fm/c.
We find that an isotropic state is achieved at about $1.0$ fm/c
in the simulation including inelastic scattering processes.
Moreover, we see that the relaxation time $\theta$ is generally time 
dependent. The two values of $\theta$ in the fits are obtained
by guessing. Actually $\theta$ can be calculated exactly, since
in order to make a local fit one should request that the time derivative 
of $F(t)$ and $Q(t)$ are equal at $t=t_0$. This leads to
\begin{equation}
\label{relax1}
\left . \dot Q(t) \right |_{t=t_0}=\left . \dot F(t) \right |_{t=t_0}
=-(Q(t_0)-Q_{eq.}) \, \frac{1}{\theta(t_0)}\,,
\end{equation}
with $Q_{eq.}=1/3$.
Changing $t_0$ to $t$ gives
\begin{equation}
\label{relax2}
\frac{\dot Q(t)}{Q_{eq.}-Q(t)}=\frac{1}{\theta(t)}.
\end{equation}
Equation (\ref{relax2}) expresses the relaxation rate $1/\theta$
of the momentum anisotropy.
In the next section we analytically separate the relaxation rate
into different terms corresponding particle diffusion and
various scattering processes, and we define the {\em transport
collision rate} which quantifies the contribution of a certain
process to momentum isotropization.

\section{Transport collision rate}
\label{sec:2}
For evaluating the momentum anisotropy at a certain space point
one has to go to its co-moving frame, in which we have
\begin{equation}
\label{aniso}
Q(t)=\left. \left < \frac{p_Z^2}{E^2} \right > \right |_{x=0}=
\frac{1}{n}\int \frac{d^3p}{(2\pi)^3}
\frac{p_Z^2}{E^2} \, f(p,x=0,t)\,.
\end{equation}
Taking the derivative in time gives
\begin{equation}
\label{dotaniso}
\dot Q(t)=\frac{1}{n}\int \frac{d^3p}{(2\pi)^3}\frac{p_Z^2}{E^2} \, 
\frac{\partial f}{\partial t}-Q(t)\frac{1}{n}\int \frac{d^3p}{(2\pi)^3}
\frac{\partial f}{\partial t}\,.
\end{equation}
We replace $\partial f/\partial t$ by
\begin{equation}
\label{boltzmann}
\frac{\partial f}{\partial t}=-\frac{\vec{p}}{E} \vec{\nabla} f +
C_{22}+C_{23}+C_{32}
\end{equation}
according to the Boltzmann equation. $C_{22}$, $C_{23}$ and $C_{32}$
denote, respectively, the collision term of $gg\to gg$, $gg\to ggg$ and
$ggg\to gg$ process. It is obvious that the contribution of different
processes to $\dot Q$ is additive. Except for a static system the 
diffusion term in Eq. (\ref{boltzmann}) generally has contribution to 
$\dot Q(t)$, which we denote by $W_{diff.}$. $C_{22}$ has no 
contribution to the second integral in Eq. (\ref{dotaniso}) due to 
particle number conservation. The same is also for the sum of $C_{23}$
and $C_{32}$ at chemical equilibrium. We rewrite Eq. (\ref{dotaniso}) to
\begin{equation}
\label{dotaniso1}
\dot Q(t)=W_{diff.}(t)+W_{22}(t)+W_{23}(t)+W_{32}(t)\,,
\end{equation}
where $W_{22}$, $W_{23}$ and $W_{32}$ correspond to the $gg\to gg$,
$gg\to ggg$ and $ggg\to gg$ process, respectively. 
According to (\ref{relax2}) we then obtain
\begin{equation}
\label{separate}
\frac{1}{\theta(t)}=R^{tr.}_{diff}(t)+R^{tr.}_{22}(t)+R^{tr.}_{23}(t)+
R^{tr.}_{32}(t)\,,
\end{equation}
where we define
\begin{equation}
\label{trate}
R^{tr.}_i(t):=\frac{W_i(t)}{Q_{eq.}-Q(t)}\,.
\end{equation}
We see that the relaxation rate of kinetic equilibration, $1/\theta$,
is separated into additive parts corresponding to particle diffusion
and collision processes. $R^{tr.}_{22}$, $R^{tr.}_{23}$ and
$R^{tr.}_{32}$ stand for the {\em transport collision rates} of the respective 
interactions and quantify their contributions to kinetic equilibration.
The extension to more than three-body processes is
straightforward because the collision term is additive.
We note that the definition of $R^{tr.}_i$ in Eq. (\ref{trate})
depends on which momentum anisotropy we are looking at. If one
defines $<|p_Z|/E>$ as the momentum anisotropy for instance, the form 
$R^{tr.}_i$ will change accordingly.

Putting the explicit expression of the collision term via
the matrix element of transition into (\ref{boltzmann}), we
obtain explicit expressions for $W_i$, which are summarized in
the following:
\begin{eqnarray}
\label{w22}
W_{22}(t)=&& n <v_{rel} \, \tilde \sigma_{22}>_2
-n \left < v_{rel} \frac{p^2_{1Z}}{E^2_1}\, \sigma_{22} \right >_2\,, \\
\label{w23}
W_{23}(t)=&& \frac{3}{2} \, n <v_{rel} \, \tilde \sigma_{23}>_2
-n \left < v_{rel} \frac{p^2_{1Z}}{E^2_1}\, \sigma_{23} \right >_2
\nonumber \\
&& -\frac{1}{2}\, Q(t)\, n <v_{rel} \,\sigma_{23}>_2\,, \\
\label{w32}
W_{32}(t)=&& \frac{1}{3} n^2 \left < \frac{\tilde I_{32}}{8E_1E_2E_3}
\right >_3-\frac{1}{2} n^2 \left < \frac{p^2_{1Z}}{E^2_1}
\frac{I_{32}}{8E_1E_2E_3} \right >_3 \nonumber \\
&& +\frac{1}{6}\, Q(t)\, n^2  \left < \frac{I_{32}}{8E_1E_2E_3}\right >_3\,,
\end{eqnarray}
where
\begin{eqnarray}
\tilde \sigma_{22}:=&&\frac{1}{2s} \frac{1}{2!} \int d\Gamma^{'}_1
d\Gamma^{'}_2 \, \frac{p^{'2}_{1Z}}{E^{'2}_1}\,
|{\cal M}_{12\to 1^{'}2^{'}}|^2 \times \nonumber \\
&& \times \, (2\pi)^4 \delta^{(4)}(p_1+p_2-p^{'}_1-p^{'}_2)\\
\tilde \sigma_{23}:=&&\frac{1}{2s} \frac{1}{3!} 
\int d\Gamma^{'}_1 d\Gamma^{'}_2 d\Gamma^{'}_3
\, \frac{p^{'2}_{1Z}}{E^{'2}_1}\,
|{\cal M}_{12\to 1^{'}2^{'}3^{'}}|^2 \times \nonumber \\
&& \times \, (2\pi)^4 \delta^{(4)}(p_1+p_2-p^{'}_1-p^{'}_2-p^{'}_3)\\
I_{32}:=&&\frac{1}{2!} \int d\Gamma^{'}_1 d\Gamma^{'}_2
\, |{\cal M}_{123\to 1^{'}2^{'}}|^2 \times \nonumber \\
&& \times \, (2\pi)^4 \delta^{(4)}(p_1+p_2+p_3-p^{'}_1-p^{'}_2)\\
\tilde I_{32}:=&&\frac{1}{2!} \int d\Gamma^{'}_1 d\Gamma^{'}_2
\, \frac{p^{'2}_{1Z}}{E^{'2}_1}\,
|{\cal M}_{123\to 1^{'}2^{'}}|^2 \times \nonumber \\
&& \times \, (2\pi)^4 \delta^{(4)}(p_1+p_2+p_3-p^{'}_1-p^{'}_2)
\label{css}
\end{eqnarray}
with $d\Gamma_i=d^3p_i/(2\pi)^3 2E_i$ for short. $\sigma_{22}$ and
$\sigma_{23}$ denote the standard pQCD cross section of
the $gg\to gg$ and $gg\to ggg$ process, respectively.
$v_{rel}=s/2E_1E_2$ is the relative velocity. $< >_2$ and $< >_3$
symbolize, respectively, an ensemble average over pairs and triplets
of incoming particles. In the parton cascade simulations 
$f(p,x,t)\approx\sum_i \delta^{(3)}(p-p_i)\delta^{(3)}(x-x_i(t))$,
and we can approximately evaluate the averages $< >_2$ and $< >_3$
in local cells which have small volume, but a sufficient number of
(test) particles to achieve high statistics. 

The expression of $W_i$s in (\ref{w22}), (\ref{w23}) and (\ref{w32})
indicates the difference of the gain and loss in the momentum
isotropization within one collision. The influence of the 
distribution of the collision angle on momentum
isotropization is implicitly contained. However, the expression of
the transport collision rate $R^{tr.}_i$ is clearly different from 
$n<v_{rel} \sigma^{tr.}_i>_2$ by the formula (\ref{tcs}). 
(The index $i$ denotes $22$ or $23$.) Only in the special case that 
all particles are moving along the $Z-$axis (irrespective of $\pm$ sign) and
have the same energy $E$,
\begin{equation}
\label{special}
f(p,x,t) \propto \delta(p_X)\delta(p_Y)\delta(p_Z-E)+
\delta(p_X)\delta(p_Y)\delta(p_Z+E)\,,
\end{equation}
the lab frame is the CM system for every colliding pair.
In this case  $p^{'2}_{1Z}/E^{'2}_1=\cos^2\theta^*$ and
\begin{equation}
\label{old}
R^{tr.}_i \sim n<v_{rel} \sigma^{tr.}_i>_2 \,.
\end{equation}
This result does not depend on the chosen direction of initial
momentum. The only necessary conditions are that all particles move
along the same direction and have the same energy.
If we define the momentum anisotropy as $<|p_Z|/E>$, $R^{tr.}_i$
maintains its form (\ref{old}), but $\sigma^{tr.}_i$ is changed to
\begin{equation}
\sigma^{tr.}_i=\int d\theta^* \frac{d\sigma_i}{d\theta^*}
(1-\cos\theta^*) \,.
\end{equation}

\begin{figure}[ht]
\begin{center}
\resizebox{0.41\textwidth}{!}{
  \includegraphics{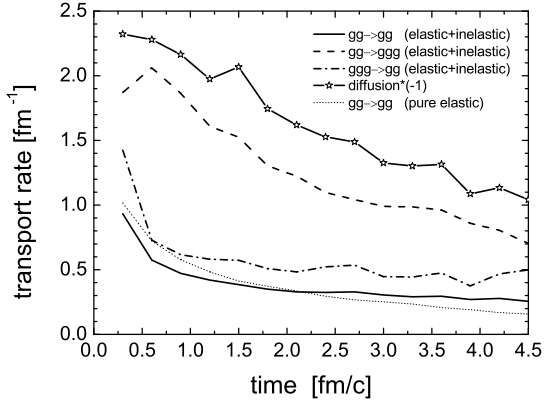}
}
\end{center}
\caption{Transport rate.}
\label{fig:2}
\end{figure}

Figure \ref{fig:2} shows the transport rates $R^{tr.}_i$ obtained 
from simulations employing the parton cascade.
The solid, dashed and dash-dotted curves depict, respectively,
the transport collision rates $R^{tr.}_{22}$, $R^{tr.}_{23}$
and $R^{tr.}_{32}$ calculated in the simulation with both elastic 
and inelastic collisions. We realize the dominance of inelastic 
collisions in kinetic equilibration by computing the ratio 
$(R^{tr.}_{23}+R^{tr.}_{32})/R^{tr.}_{22}\approx 5$.
The thin dotted curve in Fig. \ref{fig:2} presents $R^{tr.}_{22}$ in 
the simulation with pure elastic processes. When comparing this
with the solid curve one cannot realize much difference.

The contribution of particle diffusion to kinetic equilibration,
$R^{tr.}_{diff.}$, calculated in the simulation with both elastic
and inelastic processes, is depicted in Fig. \ref{fig:2} by the
symbols multiplied by $-1$. $R^{tr.}_{diff.}$, which is expressed by
\begin{equation}
\label{diff}
R^{tr.}_{diff.}(t)=\frac{1}{Q_{eq.}-Q(t)}
\frac{1}{n} \int \frac{d^3p}{(2\pi)^3}\frac{\vec p}{E} \, \cdot \vec \nabla f
\, \left ( Q(t)-\frac{p_Z^2}{E^2} \right )\,,
\end{equation}
is not computed at a certain time as performed for the transport collision 
rates, because of the inaccurate extraction of $\vec \nabla f$. The 
diffusion rate is obtained by explicit counting of the particles which come 
in as well as go out of the central region within a time interval.
Although the extraction of $R^{tr.}_{diff.}$ has a much larger 
statistical fluctuation, the sum 
$R^{tr.}_{diff.}+R^{tr.}_{22}+R^{tr.}_{23}+R^{tr.}_{32}$ gives
a value consistent with the relaxation rate $1/\theta(t)$ as one would
calculate via $\dot Q(t)/(Q_{eq.}-Q(t))$ directly from Fig. \ref{fig:1}.

We also see that $R^{tr.}_{diff.}$ has a negative contribution to 
momentum isotropization and is quite large. We did not plot
$-R^{tr.}_{diff.}$ from the simulation with pure elastic processes.
It is slightly smaller than $R^{tr.}_{22}$ (thin dotted curve). We
see that there is a big difference in the diffusion rate in both
simulations. To understand this we assume Bjorken's space-time
picture of central ultrarelativistic heavy-ion collisions \cite{bjork}
and use the relation derived by Baym \cite{baym}:
\begin{equation}
\label{baymrelat}
\frac{\vec p}{E} \cdot \vec \nabla f \approx 
\frac{p_Z}{E} \frac{\partial f}{\partial Z}
=-\frac{p_Z}{t}\frac{\partial f}{\partial p_Z}\,.
\end{equation}
Inserting Eq. (\ref{baymrelat}) into Eq. (\ref{diff}) and
calculating partial integrals give
\begin{equation}
\label{diff-2}
R^{tr.}_{diff.}(t) \approx \frac{-2}{(Q_{eq.}-Q(t))\,t}\,
\left ( Q(t)- \left < \frac{p_Z^4}{E^4} \right >(t) \right )\,.
\end{equation}
The formula confirms that $R^{tr.}_{diff.}$ is always negative. 
Using the approximation $<p_Z^4/E^4>\approx Q^2$ one can also realize
that the larger is Q, the larger $-R^{tr.}_{diff.}$ is.

\begin{figure}[t]
\begin{center}
\resizebox{0.41\textwidth}{!}{
  \includegraphics{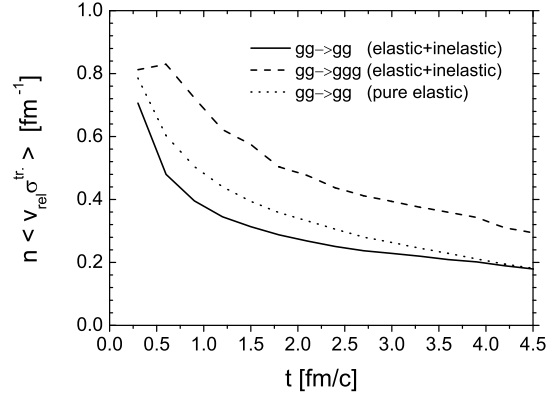} 
}
\end{center}
\caption{``Transport collision rate'' in the concept of transport cross
section.}
\label{fig:3}
\end{figure}

We have seen that only in an extreme case the transport collision rate
can be reduced to a formula directly proportional to
the transport cross section: $R^{tr.} \sim n<v_{rel}\,\sigma^{tr.}>_2$.
It is interesting to know how the calculated transport collision
rates differ from $n<v_{rel}\,\sigma^{tr.}>_2$. Such a comparison
is necessary for understanding why the concept of transport cross
section cannot explain the strong effect when including inelastic
scattering processes. In Fig. \ref{fig:3} we depict
$n<v_{rel}\,\sigma^{tr.}>_2$ calculated from the cascade simulations.
At first we look at the results in the simulation with both elastic
and inelastic collisions and calculate the inelastic to elastic ratio
(dashed versus solid curve). It is almost a constant around $1.7$ in
time, which indicates the dominance of the $gg\to ggg$ scattering
processes in kinetic equilibration. Comparing Fig. \ref{fig:3} to
Fig. \ref{fig:2} we realize that the results concerning elastic
scatterings have no strong difference. On the contrary, $R^{tr.}_{23}$
is a factor of $2.5$ larger than 
$n<v_{rel}\,\sigma^{tr.}_{23}>_2$. The pQCD $gg \to ggg$ process
is much more efficient for kinetic equilibration than one would expect
via the transport cross section. 

\section{Bremsstrahlung process and the LPM effect}
\label{sec:3}
It is intuitively clear that a $2\to 3$ process will bring one more
particle towards isotropy than a $2\to 2$ process. The kinematic
factor should be $3/2$ and appears in Eq. (\ref{w23}) when assuming
the decompositions
\begin{equation}
\left < v_{rel} \frac{p^2_{1Z}}{E^2_1}\, \sigma \right >_2 \,\approx
Q(t)<v_{rel}\, \sigma>_2\,.
\end{equation}
Analogously we compare $W_{23}$ to $W_{32}$. The sum of 
the last term in Eq. (\ref{w23}) and Eq. (\ref{w32}) comes from 
the second term in Eq. (\ref{dotaniso}) with $C_{23}+C_{32}$ instead of
$\partial f/\partial t$ and should be zero at chemical equilibrium:
we obtain
\begin{equation}
\label{chemeq}
n<v_{rel}\,\sigma_{23}>_2=\frac{1}{3}n^2
\left < \frac{I_{32}}{8E_1E_2E_3} \right >_3\,,
\end{equation}
or equivalently $R_{23}=\frac{2}{3}R_{32}$. Assuming further the decomposition
\begin{equation}
\left <\frac{p^2_{1Z}}{E^2_1}\frac{I_{32}}{8E_1E_2E_3} \right >_3
\,\approx Q(t) \left <\frac{I_{32}}{8E_1E_2E_3} \right >_3\,,
\end{equation}
we have
\begin{eqnarray*}
W_{23}(t)&\approx& \frac{3}{2} ( n <v_{rel} \, \tilde \sigma_{23}>_2
-Q(t)\, n <v_{rel} \,\sigma_{23}>_2 )\,, \\
W_{32}(t)&\approx& \frac{1}{3} n^2 \left <\frac{\tilde I_{32}}{8E_1E_2E_3}
\right >_3-Q(t)\frac{1}{3} n^2  \left <\frac{I_{32}}{8E_1E_2E_3} \right >_3.
\end{eqnarray*}
These approximate expansions together with Eq. (\ref{chemeq}) lead to 
$W_{23}\approx \frac{3}{2} W_{32}$ and
$R^{tr.}_{23}\approx \frac{3}{2} R^{tr.}_{32}$ at chemical equilibrium.
Alone due to the kinematic reason a $2\to 3$ process is $50\%$ more
efficient in kinetic equilibration than a $2\to 2$ or a $3\to 2$ process,
when $\sigma_{22}=\sigma_{23}$ and $\tilde\sigma_{22}=\tilde\sigma_{23}$. 

The LPM effect stems from the interference of the radiated gluons
(originally photons in the QED medium) by multiple scattering of a parton
though a medium. This is a coherent effect which leads to suppression
of radiation of gluons with certain modes $(w,\vec k)$. $w$ and $\vec k$
denote energy and momentum of a gluon respectively. Heuristically there
is no suppression for gluons with a {\it formation time} $\tau=w/k_T^2$
smaller than the mean free path. This is called the 
Bethe-Heitler limit, where the gluon radiations induced at a different
space-time point in the course of the propagation of a parton can
be considered as independent events. These events within the 
Bethe-Heitler regime have been included in the parton cascade
calculations. Radiation of other gluon modes with coherent suppression
completely dropped out, which is indicated by the $\Theta$-function
in the matrix element (\ref{m23}). The inclusion of those radiations
would speed up thermalization. How to implement the coherent effect
into a transport model solving the Boltzmann equation is a challenge.

The $\Theta$-function in the matrix element (\ref{m23}) results in
a cut-off for $k_T$, the transverse momentum of a radiated gluon,
$k_T>1/\Lambda_g$, where $\Lambda_g$ is the mean free path of a gluon.
A higher value of the cut-off will decrease the total cross section of
a $gg\to ggg$ collision on the one hand and make the collision angles
large on the other hand. The latter leads to a large efficiency
for momentum isotropization. Varying the cut-off downwards to a
smaller value one would enter into the LPM suppressed regime.
Although parton cascade calculations set up with smaller cut-offs cannot
completely take the LPM effect into account, one can roughly
estimate the contribution of the coherent effect to kinetic
equilibration. Such calculations will be done in a subsequent paper.

\section{Conclusion}
\label{conclu}
Employing the parton cascade we have investigated the importance
of including pQCD Bremsstrahlung processes to thermalization.
The question addressed is how to understand the observed fast
equilibration in theoretical terms. The special emphasis is put on 
expressing the transport collision rate in a correct manner. 
The concept of transport cross 
section only gives a qualitative understanding of the dominant 
contribution of large-angle scatterings to momentum isotropization
but not a correct way to manifest the various contributions. In case
we are studying parton thermalization in a central Au+Au collision at
RHIC energy the old concept of the transport cross sections would
strongly underestimate the contribution of $gg\to ggg$ to kinetic 
equilibration. The correct results showed that the inclusion of pQCD 
Bremsstrahlung processes increases the efficiency by a factor of $5$ 
for thermalization. The large efficiency stems partly from the increase 
of particle number in the final state of $gg\to ggg$ collisions, but 
mainly from the almost isotropic angular distribution in Bremsstrahlung 
processes due to the effective implementation of 
LPM suppression. The detailed understanding of
the latter has to be developed in future investigations.

%

\begin{thebibliography}{}
%
%
\bibitem{Guyl}
Miklos Gyulassy and Larry McLerran, Nucl.Phys.A \textbf{750}, 
(2005) 30-63.
\bibitem{Xu1}
Zhe Xu and Carsten Greiner, Phys.Rev.C \textbf{71}, (2005) 064901.
\bibitem{Xu2}
Zhe Xu and Carsten Greiner, Proceedings of Quark Matter 2006,
hep-ph/0509324.
\bibitem{serreau}
J. Serreau and D. Schiff, J. High Energy Phys. \textbf{0111}, (2001) 039.
\bibitem{dg}
P. Danielewicz and M. Gyulassy, Phys.Rev.D \textbf{31}, (1985) 53.
\bibitem{molnar}
D. Molnar, Proceedings of Quark Matter 1999,
Nucl.Phys.A \textbf{661}, (1999) 205-260.
\bibitem{gb}
J.F. Gunion and G. Bertsch, Phys.Rev.D \textbf{25}, (1982) 746.
\bibitem{biro}
T.S. Biro et al, Phys.Rev.C \textbf{48}, (1993) 1275.
\bibitem{wong}
S.M.H. Wong, Nucl.Phys.A \textbf{607}, (1996) 442.
\bibitem{bjork}
J.D. Bjorken, Phys.Rev.D \textbf{27}, (1983) 140.
\bibitem{baym}
G. Baym, Phys.Lett.B \textbf{138}, (1984) 18.
\end{thebibliography}
%

\end{document}